\documentclass[12pt,titlepage]{article}
\usepackage{amsfonts}
\usepackage{amsmath}
\usepackage{amssymb}

\begin{document}

\title{Spinless bosons embedded in a vector Duffin-Kemmer-Petiau oscillator}
\author{L.B. Castro\thanks{%
benito@feg.unesp.br }, A.S. de Castro\thanks{%
castro@pq.cnpq.br.} \\
\\
UNESP - Campus de Guaratinguet\'{a}\\
Departamento de F\'{\i}sica e Qu\'{\i}mica\\
12516-410 Guaratinguet\'{a} SP - Brazil}
\date{}
\maketitle

\begin{abstract}
Some properties of minimal and nonminimal vector interactions in the
Duffin-Kemmer-Petiau (DKP) formalism are discussed. The conservation of the
total angular momentum for spherically symmetric nonminimal potentials is
derived from its commutation properties with each term of the DKP equation
and the proper boundary conditions on the spinors are imposed. It is shown
that the space component of the nonminimal vector potential plays a crucial
role for the confinement of bosons. The exact solutions for the vector DKP
oscillator (nonminimal vector coupling with a linear potential which
exhibits an equally spaced energy spectrum in the weak-coupling limit) for
spin-0 bosons are presented in a closed form and it is shown that the
spectrum exhibits an accidental degeneracy.\newline
\newline
\newline


\end{abstract}

\section{Introduction}

The first-order Duffin-Kemmer-Petiau (DKP) formalism \cite{pet}-\cite{kem}
describes spin-0 and spin-1 particles. The DKP equation for a free boson is
given by \cite{kem} $\left( \beta ^{\mu }p_{\mu }-m\right) \psi =0$ (with
units in which $\hbar =c=1$), \noindent where the four beta matrices satisfy
the algebra $\beta ^{\mu }\beta ^{\nu }\beta ^{\lambda }+\beta ^{\lambda
}\beta ^{\nu }\beta ^{\mu }=g^{\mu \nu }\beta ^{\lambda }+g^{\lambda \nu
}\beta ^{\mu }$\noindent\ and the metric tensor is $g^{\mu \nu }=\,$diag$%
\,(1,-1,-1,-1)$. The algebra expressed by those matrices generates a set of
126 independent matrices whose irreducible representations are a trivial
representation, a five-dimensional representation describing the spin-0
particles and a ten-dimensional representation associated to spin-1
particles. A well-known conserved four-current is given by $j^{\mu }=\bar{%
\psi}\beta ^{\mu }\psi $\noindent $/2$, where the adjoint spinor $\bar{\psi}$
is given by $\bar{\psi}=\psi ^{\dagger }\eta ^{0}$ with $\eta ^{\mu }=2\beta
^{\mu }\beta ^{\mu }-g^{\mu \mu }$\textrm{\ }in such a way that $\left( \eta
^{0}\beta ^{\mu }\right) ^{\dagger }=\eta ^{0}\beta ^{\mu }$ (the matrices $%
\beta ^{\mu }$ are Hermitian with respect to $\eta ^{0}$). Despite the
similarity to the Dirac equation, the DKP equation involves singular
matrices, the time component of $j^{\mu }$ is not positive definite and the
case of massless bosons can not be obtained by a limiting process.
Nevertheless, the matrices $\beta ^{\mu }$ plus the unit operator generate a
ring consistent with integer-spin algebra \cite{nie} and $j^{0}$ may be
interpreted as a charge density. The factor 1/2 multiplying $\bar{\psi}\beta
^{\mu }\psi $, of no importance regarding the conservation law, is in order
to hand over a charge density conformable to that one used in the
Klein-Gordon theory and its nonrelativistic limit (see e.g. \cite{puk}).
Then the normalization condition $\int d\tau \,j^{0}=\pm 1$ can be expressed
as $\int d\tau \,\bar{\psi}\beta ^{0}\psi =\pm 2$, where the plus (minus)
sign must be used for a positive (negative) charge.

The DKP formalism has been used to analyze relativistic interactions of
spin-0 and spin-1 hadrons with nuclei. A number of different couplings in
the DKP formalism, with scalar and vector couplings in analogy with the
Dirac phenomenology for proton-nucleus scattering, has been employed in the
phenomenological treatment of the elastic meson-nucleus scattering at medium
energies with a better agreement to the experimental data when compared to
the Klein-Gordon and Proca based formalisms \cite{cla1}-\cite{cla2}.
Recently, there has been an increasing interest on the so-called DKP
oscillator \cite{deb}-\cite{boucheto}. That system is a kind of tensor
coupling with a linear potential which leads to the harmonic oscillator
problem in the weak-coupling limit. A nonminimal vector potential, added by
other kinds of Lorentz structures, has already been used successfully in a
phenomenological context for describing the scattering of mesons by nuclei
\cite{cla1}-\cite{kal}, \cite{kur1}, \cite{cla2}, and a sort of vector DKP
oscillator (nonminimal vector coupling with a linear potential \cite{kuli},
\cite{tati1}) has also been an item of recent investigation. Vector DKP
oscillator is the name given to the system with a Lorentz vector coupling
which exhibits an equally spaced energy spectrum in the weak-coupling limit.
The name distinguishes from that system called DKP oscillator with Lorentz
tensor couplings of Ref. \cite{deb}-\cite{boucheto}. The nonminimal vector
coupling with square step \cite{ccc2} and smooth step potentials \cite{ccc4}
have also appeared in the literature.

The one-dimensional vector DKP oscillator was treated in Ref. \cite{tati1}
but we show in this paper that the three-dimensional case has some very
special features such as the question of conservation of the total angular
momentum $\overrightarrow{J}$, boundary conditions on the spinor and
degeneracy of the spectrum. The conservation of $\overrightarrow{J}$ is
derived from its commutation properties with each term of the DKP equation.
The proper boundary condition at the origin follows from the absence of
Dirac delta potentials, avoiding in this manner to recourse to plausibility
arguments regarding the self-adjointness of the momentum and the finiteness
of the kinetic energy, as done by Greiner \cite{gre} in the case of the
nonrelativistic harmonic oscillator. The exact solutions are presented in a
closed form and the spectrum presents, beyond the essential degeneracy
omnipresent for any central force field, an accidental degeneracy.

\section{Vector interactions in the DKP equation}

With the introduction of interactions, the DKP equation can be written as%
\begin{equation}
\left( \beta ^{\mu }p_{\mu }-m-V\right) \psi =0  \label{dkp2}
\end{equation}%
where the more general potential matrix $V$ is written in terms of 25 (100)
linearly independent matrices pertinent to the five(ten)-dimensional
irreducible representation associated to the scalar (vector) sector. In the
presence of interactions $j^{\mu }$ satisfies the equation%
\begin{equation}
\partial _{\mu }j^{\mu }+\frac{i}{2}\bar{\psi}\left( V-\eta ^{0}V^{\dagger
}\eta ^{0}\right) \psi =0  \label{corrent2}
\end{equation}%
Thus, if $V$ is Hermitian with respect to $\eta ^{0}$ then the four-current
will be conserved. The potential matrix $V$ can be written in terms of
well-defined Lorentz structures. For the spin-0 sector there are two scalar,
two vector and two tensor terms \cite{gue}, whereas for the spin-1 sector
there are two scalar, two vector, a pseudoscalar, two pseudovector and eight
tensor terms \cite{vij}. The tensor terms have been avoided in applications
because they furnish noncausal effects \cite{gue}-\cite{vij}. Considering
only the vector terms, $V$ is in the form%
\begin{equation}
V=\beta ^{\mu }A_{\mu }^{\left( 1\right) }+i[P,\beta ^{\mu }]A_{\mu
}^{\left( 2\right) }  \label{pot}
\end{equation}%
where $P$ is a projection operator ($P^{2}=P$ and $P^{\dagger }=P$) in such
a way that $\bar{\psi}P\psi $ behaves as a scalar and $\bar{\psi}[P,\beta
^{\mu }]\psi $ behaves like a vector. $A_{\mu }^{\left( 1\right) }$ and $%
A_{\mu }^{\left( 2\right) }$ are the four-vector potential functions. Notice
that the vector potential $A_{\mu }^{\left( 1\right) }$ is minimally coupled
but not $A_{\mu }^{\left( 2\right) }$. One very important point to note is
that this matrix potential leads to a conserved four-current but the same
does not happen if instead of $i[P,\beta ^{\mu }]$ one uses either $P\beta
^{\mu }$ or $\beta ^{\mu }P$, as in \cite{cla1}-\cite{kal}, \cite{kur1},
\cite{cla2}, \cite{nab}). As a matter of fact, in Ref. \cite{cla1} is
mentioned that $P\beta ^{\mu }$ and $\beta ^{\mu }P$ produce identical
results.

The DKP equation is invariant under the parity operation, i.e. when $%
\overrightarrow{r}\rightarrow -\overrightarrow{r}$, if $\overrightarrow{A}%
^{\left( 1\right) }$ and $\overrightarrow{A}^{\left( 2\right) }$ change
sign, whereas $A_{0}^{\left( 1\right) }$ and $A_{0}^{\left( 2\right) }$
remain the same. This is because the parity operator is $\mathcal{P}=\exp
(i\delta _{P})P_{0}\eta ^{0}$, where $\delta _{P}$ is a constant phase and $%
P_{0}$ changes $\overrightarrow{r}$ into $-\overrightarrow{r}$. Because this
unitary operator anticommutes with $\overrightarrow{\beta }$ and $[P,%
\overrightarrow{\beta }]$, they change sign under a parity transformation,
whereas $\beta ^{0}$ and $[P,\beta ^{0}]$, which commute with $\eta ^{0}$,
remain the same. Since $\delta _{P}=0$ or $\delta _{P}=\pi $, the spinor
components have definite parities. The charge-conjugation operation changes
the sign of the minimal interaction potential, i.e. changes the sign of $%
A_{\mu }^{\left( 1\right) }$. This can be accomplished by the transformation
$\psi \rightarrow \psi _{c}=\mathcal{C}\psi =CK\psi $, where $K$ denotes the
complex conjugation and $C$ is a unitary matrix such that $C\beta ^{\mu
}=-\beta ^{\mu }C$. The matrix that satisfies this relation is $C=\exp
\left( i\delta _{C}\right) \eta ^{0}\eta ^{1}$. The phase factor $\exp
\left( i\delta _{C}\right) $ is equal to $\pm 1$, thus $E\rightarrow -E$.
Note also that $j^{\mu }\rightarrow -j^{\mu }$, as should be expected for a
charge current. Meanwhile $C$ anticommutes with $[P,\beta ^{\mu }]$ and the
charge-conjugation operation entails no change on $A_{\mu }^{\left( 2\right)
}$. The invariance of the nonminimal vector potential under charge
conjugation means that it does not couple to the charge of the boson. In
other words, $A_{\mu }^{\left( 2\right) }$ does not distinguish particles
from antiparticles. Hence, whether one considers spin-0 or spin-1 bosons,
this sort of interaction can not exhibit Klein's paradox.

For massive spinless bosons the projection operator is given by \cite{gue}
\begin{equation}
P=\frac{1}{3}\left( \beta ^{\mu }\beta _{\mu }-1\right)  \label{pro}
\end{equation}%
\noindent Defining $P^{\mu }=P\beta ^{\mu }$ and $^{\mu }P=\beta ^{\mu }P$,
one can obtain the follow relations \cite{FIS}
\begin{equation*}
\beta ^{\mu }=P^{\mu }+\,^{\mu }P,\,\,\,\,\,\,P^{\mu }\beta ^{\nu }=Pg^{\mu
\nu }
\end{equation*}%
\begin{equation}
(P^{\mu })P=P(^{\mu }P)=0,\,\,\,\,\,\,(P^{\mu })(P^{\nu })=(^{\mu }P)(^{\nu
}P)=0  \label{rela}
\end{equation}%
Applying $P$ and $P^{\nu }$ to the DKP equation and using the relations (\ref%
{rela}), we have
\begin{equation}
i\left( D_{\mu }-A_{\mu }^{(2)}\right) (P^{\mu }\psi )=m(P\psi )  \label{p1}
\end{equation}%
\noindent and
\begin{equation}
i(D_{\mu }+A_{\mu }^{(2)})(P\psi )=m(P_{\mu }\psi )  \label{p2}
\end{equation}%
\noindent respectively. Here, $D_{\mu }=\partial _{\mu }+iA_{\mu }^{(1)}$.
Combining these results we obtain
\begin{equation}
\left[ D^{\mu }D_{\mu }+m^{2}+(\partial ^{\mu }A_{\mu }^{(2)})-\left(
A^{(2)}\right) _{\mu }\left( A^{(2)}\right) ^{\mu }\right] (P\psi )=0
\label{dkpp}
\end{equation}%
\noindent On the other hand, by using (\ref{rela}) $j^{\mu }$ can be written
as

\begin{equation}
j^{\mu }=-\frac{1}{m}\,\mathrm{Im}\left[ (P\psi )^{\dagger }D^{\mu }(P\psi )%
\right]  \label{ju}
\end{equation}

\noindent One sees that $A_{\mu }^{(2)}$ does not intervene explicitly in
the current and, in the absence of the nonminimal potential, (\ref{dkpp})
reduces to the Klein-Gordon equation in the presence of a minimally coupled
potential and that all elements of the column matrix $P\psi $ are scalar
fields of mass $m$. It is instructive to note that the form of the two
distinct vector couplings in the generalized Klein-Gordon equation has
become obvious because the interaction operates under the umbrella of the
DKP theory. Otherwise, only the minimal vector coupling could be obtained by
applying the minimal substitution $\partial _{\mu }\rightarrow \partial
_{\mu }+iA_{\mu }^{(1)}$ to the free Klein-Gordon equation.

\section{The nonminimal vector interaction}

In this stage, we concentrate our efforts in the nonminimal vector potential
$A_{\mu }^{(2)}=A_{\mu }$ and use the representation for the $\beta ^{\mu }$%
\ matrices given by \cite{deb}, \cite{ned1}%
\begin{equation}
\beta ^{0}=%
\begin{pmatrix}
\theta & \overline{0} \\
\overline{0}^{T} & \mathbf{0}%
\end{pmatrix}%
,\quad \quad \overrightarrow{\beta }=%
\begin{pmatrix}
\widetilde{0} & \overrightarrow{\rho } \\
-\overrightarrow{\rho }^{\,T} & \mathbf{0}%
\end{pmatrix}%
\end{equation}%
\noindent where%
\begin{eqnarray}
\ \theta &=&%
\begin{pmatrix}
0 & 1 \\
1 & 0%
\end{pmatrix}%
,\quad \quad \rho ^{1}=%
\begin{pmatrix}
-1 & 0 & 0 \\
0 & 0 & 0%
\end{pmatrix}
\notag \\
&& \\
\rho ^{2} &=&%
\begin{pmatrix}
0 & -1 & 0 \\
0 & 0 & 0%
\end{pmatrix}%
,\quad \quad \rho ^{3}=%
\begin{pmatrix}
0 & 0 & -1 \\
0 & 0 & 0%
\end{pmatrix}
\notag
\end{eqnarray}%
\noindent $\overline{0}$, $\widetilde{0}$ and $\mathbf{0}$ are 2$\times $3, 2%
$\times $2 and 3$\times $3 zero matrices, respectively, while the
superscript T designates matrix transposition. The five-component spinor can
be written as $\psi ^{T}=\left( \psi _{1},\ldots ,\psi _{5}\right) $. With
this representation the projection operator is $P=\mathrm{diag}(1,0,0,0,0)$.
In this case $P$ picks out the first component of the DKP spinor.

If the terms in the potential $A^{\mu }$ are time-independent one can write $%
\psi (\overrightarrow{r},t)=\phi (\overrightarrow{r})\exp (-iEt)$, where $E$
is the energy of the boson, in such a way that the time-independent DKP
equation becomes%
\begin{equation}
\left[ \beta ^{0}E+i\beta ^{i}\partial _{i}-\left( m+i[P,\beta ^{\mu
}]A_{\mu }\right) \right] \phi =0  \label{DKP10}
\end{equation}%
In this case \ $j^{\mu }=\bar{\phi}\beta ^{\mu }\phi /2$ does not depend on
time, so that the spinor $\phi $ describes a stationary state. In the
time-independent case (\ref{p2}) becomes%
\begin{equation}
\phi _{2}=\frac{1}{m}\left( E+iA_{0}\right) \,\phi _{1}  \label{dkp3a}
\end{equation}

\begin{equation}
\overrightarrow{\zeta }=\left( \overrightarrow{\nabla }-\overrightarrow{A}%
\right) \phi _{1}  \label{dkp3}
\end{equation}%
where%
\begin{equation}
\overrightarrow{\zeta }=\frac{m}{i}(\phi _{3},\phi _{4},\phi _{5})^{T}
\end{equation}%
and (\ref{dkpp}) furnishes
\begin{equation}
\left( -\nabla ^{2}+\overrightarrow{\nabla }\cdot \overrightarrow{A}+%
\overrightarrow{A}^{2}\right) \phi _{1}=k^{2}\phi _{1}
\end{equation}%
where%
\begin{equation}
k^{2}=E^{2}-m^{2}+A_{0}^{2}  \label{k}
\end{equation}

\noindent Meanwhile,
\begin{equation}
j^{0}=\frac{E}{m}\,|\phi _{1}|^{2},\quad \overrightarrow{j}=\frac{1}{m}\,%
\mathrm{Im}\left( \phi _{1}^{\ast }\,\overrightarrow{\nabla }\phi _{1}\right)
\end{equation}%
\noindent

If we consider spherically symmetric potentials
\begin{equation}
A^{\mu }(\overrightarrow{r})=\left( A_{0}(r),A_{r}(r)\hat{r}\right)
\label{pote}
\end{equation}%
\noindent then the DKP equation permits the factorization
\begin{equation}
\phi _{1}(\overrightarrow{r})=\frac{u_{\kappa }(r)}{r}\,Y_{lm_{l}}(\theta
,\varphi )  \label{sp}
\end{equation}%
\noindent where $Y_{lm_{l}}$ is the usual spherical harmonic, with $%
l=0,1,2,\ldots $, $m_{l}=-l,-l+1,\ldots ,l$, $\int d\Omega
\,Y_{lm_{l}}^{\ast }Y_{l^{\prime }m_{l^{\prime }}}=\delta _{ll^{\prime
}}\delta _{m_{l}m_{l^{\prime }}}$ and $\kappa $ stands for all quantum
numbers which may be necessary to characterize $\phi _{1}$. For $r\neq 0$
the radial function $u$ obeys the radial equation
\begin{equation}
\frac{d^{2}u}{dr^{2}}+\left[ k^{2}-2\,\frac{A_{r}}{r}-\frac{dA_{r}}{dr}-%
\frac{l(l+1)}{r^{2}}-A_{r}{}^{2}\right] u=0  \label{21}
\end{equation}%
\noindent and because $\nabla ^{2}\left( 1/r\right) =-4\pi \delta (%
\overrightarrow{r})$, unless the potentials contain a delta function at the
origin, one must impose the homogeneous Dirichlet condition $u\left(
0\right) =0$ \cite{sha}. Furthermore, from the normalization condition $\int
d\tau \,j^{0}=\pm 1$ one sees that $u$ must be normalized according to%
\begin{equation}
\frac{E}{m}\int_{0}^{\infty }dr\,|u|^{2}\,=\pm 1
\end{equation}%
Therefore, for motion in a central field, the solution of the
three-dimensional DKP equation can be found by solving a Schr\"{o}%
dinger-like equation. The other components are obtained through of (\ref%
{dkp3a}) and (\ref{dkp3}). Note that the DKP spinor is an eigenstate of the
parity operator. This happens because $\eta ^{0}=\,$diag$\,(1,1,-1,-1,-1)$
and the parity of $\overrightarrow{\zeta }$ is opposite to that one of $\phi
_{1}$ and $\phi _{2}$. Furthermore, the spin operator $S_{k}=i\varepsilon
_{klm}\beta ^{l}\beta ^{m}$ \cite{kem} satisfies the commutation relations%
\begin{eqnarray}
\lbrack S_{k},\beta ^{0}] &=&[S_{k},[P,\beta ^{0}]]=0  \notag \\
&&  \label{opSPIN} \\
\lbrack S_{k},\beta ^{l}] &=&i\varepsilon _{klm}\beta ^{m},\quad \lbrack
S_{k},[P,\beta ^{l}]]=i\varepsilon _{klm}[P,\beta ^{m}]  \notag
\end{eqnarray}%
so that the total angular momentum $\overrightarrow{J}=\overrightarrow{L}+%
\overrightarrow{S}$ satisfies%
\begin{equation}
\lbrack \overrightarrow{J},\beta ^{\mu }p_{\mu }]=[\overrightarrow{J},\beta
^{\mu }A_{\mu }^{\left( 1\right) }]=[\overrightarrow{J},[P,\beta ^{\mu
}]A_{\mu }^{\left( 2\right) }]=\overrightarrow{0}  \label{opJ}
\end{equation}%
in such a way that the DKP spinor is also an eigenstate of $\overrightarrow{J%
}^{2}$ and $J_{3}$. Accordingly, the DKP spinor can be classified by the
parity, by the total angular momentum, and its third component, quantum
numbers. As a matter of fact,%
\begin{equation}
\overrightarrow{S}=%
\begin{pmatrix}
\widetilde{0} & \overline{0} \\
\overline{0}\,^{T} & \overrightarrow{s}%
\end{pmatrix}
\label{jotaM}
\end{equation}%
where $s_{k}$ are the 3$\times $3 spin-1 matrices $\left( s_{k}\right)
_{lm}=-i\varepsilon _{klm}$. As a result, $\overrightarrow{S}$ does not act
on the two upper components of the DKP spinor. This means that the orbital
angular momentum quantum numbers of $\phi _{1}$ and $\phi _{2}$ are good
quantum numbers. With the orbital angular momentum quantum number $l$
referring to the two upper components of the DKP spinor, as before, $%
\overrightarrow{\zeta }$ in (\ref{dkp3}) can be written as \cite{arf}
\begin{equation*}
\overrightarrow{\zeta }=\overrightarrow{Y}\!_{l,l-1,m_{l}}\sqrt{\frac{l}{2l+1%
}}\left( \frac{d}{dr}+\frac{l+1}{r}-A_{r}^{(2)}\right) \frac{u(r)}{r}
\end{equation*}

\begin{equation}
-\overrightarrow{Y}\!_{l,l+1,m_{l}}\sqrt{\frac{l+1}{2l+1}}\left( \frac{d}{dr}%
-\frac{l}{r}-A_{r}^{(2)}\right) \frac{u(r)}{r}  \label{zetaV}
\end{equation}%
In this last expression, $\overrightarrow{Y}\!_{Jlm_{J}}(\theta ,\varphi )$
are the so-called vector spherical harmonics. They result from the coupling
of the three dimensional unit vectors in spherical notation to the
eigenstates of orbital angular momentum, form a complete orthonormal set and
satisfy%
\begin{equation}
\overrightarrow{J}^{2}\overrightarrow{Y}\!_{Jlm_{J}}=J\left( J+1\right)
\overrightarrow{Y}\!_{Jlm_{J}},\quad \overrightarrow{L}^{2}\overrightarrow{Y}%
\!_{Jlm_{J}}=l\left( l+1\right) \overrightarrow{Y}\!_{Jlm_{J}},\quad J_{3}%
\overrightarrow{Y}\!_{Jlm_{J}}=m_{J}\overrightarrow{Y}\!_{Jlm_{J}}
\label{rel1}
\end{equation}%
and $\overrightarrow{Y}_{l,l\pm 1,m_{l}}$ transforms under parity as
\begin{equation}
\overrightarrow{Y}\!_{l,l\pm 1,m_{l}}(\theta -\pi ,\varphi +\pi )=\left(
-1\right) ^{l+1}\overrightarrow{Y}\!_{l,l\pm 1,m_{l}}(\theta ,\varphi )
\label{rel2}
\end{equation}%
One sees that if the two upper components of the DKP spinor are
eigenfunctions of $\overrightarrow{L}^{2}$ with an orbital angular momentum
quantum number $l$, the three lower components will be a linear
superposition of two types of eigenfunctions of \ $\overrightarrow{L}^{2}$.
One of those with orbital angular momentum quantum number $l+1$ and the
other with $l-1$. The fact that the upper and lower components of the DKP
spinor have different orbital angular momentum quantum numbers is related to
the fact that $\overrightarrow{L}$ is not a conserved quantity in the DKP
theory. Nevertheless, the orbital angular momentum quantum number of the
first component of the DKP spinor equals the total angular momentum quantum
number of the DKP spinor, as it should be since $\phi _{1}$ describes a
spinless particle. It follows that the parity of the DKP spinor is given by $%
\left( -1\right) ^{l}$.

\section{The vector DKP oscillator}

Let us consider a nonminimal vector linear potential in the form%
\begin{equation}
A_{0}^{(2)}=m^{2}\lambda _{0}r,\quad A_{r}^{(2)}=m^{2}\lambda _{r}r
\label{pot2}
\end{equation}%
where $\lambda _{0}$ and $\lambda _{r}$ are dimensionless quantities. Our
problem is to solve (\ref{21}) for $u$ and to determine the allowed energies.

One finds that $u$ obeys the second-order differential equation%
\begin{equation}
\frac{d^{2}u}{dr^{2}}+\left[ K^{2}-\lambda ^{2}r^{2}-\frac{l\left(
l+1\right) }{r^{2}}\right] u=0  \label{eqr}
\end{equation}%
where
\begin{equation}
K=\sqrt{E^{2}-m^{2}-3m^{2}\lambda _{r}},\quad \lambda =m^{2}\sqrt{\lambda
_{r}^{2}-\lambda _{0}^{2}}
\end{equation}%
With $u(0)=0$ and $\int_{0}^{\infty }dr\,|u|^{2}\,<\infty $, the solution
for (\ref{eqr}) with $K$ and $\lambda $ real is precisely the well-known
solution of the Schr\"{o}dinger equation for the three-dimensional harmonic
oscillator (see, e.g. \cite{gre}). For $\lambda =i|\lambda |$ (and $K=|K|$
or $K=i|K|$), the case of an inverted harmonic oscillator, the energy
spectrum will consist of a continuum corresponding to unbound states. We
shall limit ourselves to study the case of bound-state solutions.

The asymptotic behaviour of (\ref{eqr}) and the conditions $u(0)=0$ and $%
\int_{0}^{\infty }dr\,|u|^{2}\,<\infty $ dictate that the solution close to
the origin valid for all values of $l$ can be written as being proportional
to $r^{l+1}$, and proportional to $e^{-\lambda r^{2}/2}$ as $r\rightarrow
\infty $. It is convenient to introduce the following new variable and
parameters:%
\begin{equation}
z=\lambda r^{2},\quad a=\frac{1}{2}\left( l+\frac{3}{2}-\frac{K^{2}}{%
2\lambda }\right) ,\quad b=l+\frac{3}{2}  \label{20}
\end{equation}%
so that the solution for all $r$ can be expressed as $u(r)=r^{l+1}e^{-%
\lambda r^{2}/2}w(r)$, where $w$ is a regular solution of the confluent
hypergeometric equation (Kummer's equation) \cite{abr}

\begin{equation}
z\frac{d^{2}w}{dz^{2}}+(b-z)\frac{dw}{dz}-aw=0  \label{19}
\end{equation}

\noindent The general solution of (\ref{19}) is given by \cite{abr}
\begin{equation}
w=A\,\,M(a,b,z)+B\,z^{1-b}\,\,M(a-b+1,2-b,z)  \label{W}
\end{equation}%
with arbitrary constants $A$ and $B$. The confluent hypergeometric function
(Kummer's function) $M(a,b,z)$, or $_{1}F_{1}(a,b,z)$, is$\,$%
\begin{equation}
M(a,b,z)=\frac{\Gamma \left( b\right) }{\Gamma \left( a\right) }%
\sum_{n=0}^{\infty }\frac{\Gamma \left( a+n\right) }{\Gamma \left(
b+n\right) }\,\frac{z^{n}}{n!}  \label{m}
\end{equation}%
and $\Gamma \left( z\right) $ is the gamma function. The second term in (\ref%
{W}) has a singular point at $z=0$, so we set $B=0$. In order to furnish
normalizable $\phi _{1}$, the confluent hypergeometric function must be a
polynomial. This is because $M(a,b,\lambda r^{2})$ goes as $e^{\lambda r^{2}}
$ as $r$ goes to infinity unless the series breaks off. This demands that $%
a=-n$, where $n$ is a nonnegative integer. We put $N=2n+l$, whence the
requirement $a=-n$ implies into
\begin{equation}
|E|=m\sqrt{1+3\lambda _{r}+\left( 2N+3\right) \sqrt{\lambda _{r}^{2}-\lambda
_{0}^{2}}},\quad N=0,1,2,\ldots   \label{37}
\end{equation}%
\noindent\ Note that $M(-n,b,\lambda r^{2})$ is proportional to the
generalized Laguerre polynomial $L_{n}^{\left( b-1\right) }(\lambda r^{2})$,
a polynomial of degree $n$. Therefore, by using the normalization condition $%
\left\vert \int d\tau \,j^{0}\right\vert =|E|/m\,\int d\tau \,|\phi
_{1}|^{2}=1$, with $|E|\neq 0$, and that the generalized Laguerre polynomial
is standardized as \cite{abr}
\begin{equation}
\int\limits_{0}^{\infty }d\xi \,\xi ^{\alpha }e^{-\xi }\left[ L_{n}^{\left(
\alpha \right) }(\xi )\right] ^{2}=\frac{\Gamma (\alpha +n+1)}{n!}
\label{stand}
\end{equation}%
one determines $A$ in (\ref{W}) and obtains%
\begin{equation}
\phi _{1}=\sqrt{\frac{2m\lambda ^{l+3/2}\left( \frac{N-l}{2}\right) !}{%
|E|\,\Gamma \left( \frac{N+l+3}{2}\right) }}\,r^{l}e^{-\lambda r^{2}/2}\;L_{%
\frac{N-l}{2}}^{\left( l+1/2\right) }\left( \lambda r^{2}\right)
Y_{lm_{l}}(\theta ,\varphi )  \label{38}
\end{equation}%
Note that $l$ can take the values $0,2,...,N$ when $N$ is an even number,
and $1,3,...,N$ when $N$ is an odd number. All the energy levels with the
exception of that one for $N=0$ are degenerate. The degeneracy of the level
of energy for a given principal quantum number $N$ is given by $\left(
N+1\right) \left( N+2\right) /2$. Notice that the condition $\lambda $ $\in
\mathbb{R}
$ requires that $|\lambda _{r}|>|\lambda _{0}|$, meaning that the space
component of the potential must be stronger than its time component. There
is an infinite set of discrete energies (symmetrical about $E=0$ as it
should be since $A_{\mu }$ does not distinguish particles from
antiparticles) irrespective to the sign of $\lambda _{0}$. In general, $|E|$
is higher for $\lambda _{r}>0$ than for $\lambda _{r}<0$. It increases with
the principal quantum number and it is a monotonically decreasing function
of $\lambda _{0}$. For $\lambda _{r}<0$ and $\lambda _{0}=0$ the spectrum
acquiesces $|E|=m$ for $N=0$. In order to insure the reality of the
spectrum, the coupling constants $\lambda _{0}$ and $\lambda _{r}$ satisfy
the additional constraint $\sqrt{\lambda _{r}^{2}-\lambda _{0}^{2}}>-\left(
1+3\lambda _{r}\right) /(2N+3)$ in such a way that there can be no bound
states for $\lambda _{r}<0$ with small principal quantum numbers and $%
|\lambda _{r}|$ enough small. This means that for $\lambda _{r}<0$ and $%
|\lambda _{r}|$ enough small a number of solutions with the smallest
principal quantum numbers does not exist. For $|\lambda _{r}|\simeq |\lambda
_{0}|$ we have a very high density of very delocalized states (because $%
\lambda \simeq 0$). For $|\lambda _{r}|\gg |\lambda _{0}|$ one has that%
\begin{equation}
|E|\simeq m\sqrt{1+3\lambda _{r}+\left( 2N+3\right) |\lambda _{r}|}
\label{lim2}
\end{equation}%
so that $|E|>m$ for $\lambda _{r}>0$. Concerning $\lambda _{r}<0$, as far as
$\lambda _{r}$ decreases, the spectrum moves towards $E=0$, except for $%
\lambda _{0}=0$ which maintains $|E|\geq m$. On the other hand, in the
weak-coupling limit, $|\lambda _{r}|\ll 1$ and $|\lambda _{0}|\ll 1$, $%
|E|\simeq m$ for small quantum numbers, and (\ref{37}) becomes%
\begin{equation}
|E|\simeq m\left[ 1+\frac{3}{2}\,\lambda _{r}+\left( N+\frac{3}{2}\right)
\sqrt{\lambda _{r}^{2}-\lambda _{0}^{2}}\right]   \label{nr}
\end{equation}%
Because of this equally spaced energy spectrum, it can be said that the
linear potential given by (\ref{pot2}) describes a genuine vector DKP
oscillator. It is obvious that, despite the effective harmonic oscillator
potential appearing in (\ref{eqr}) and the spectrum given by (\ref{nr}), in
a nonrelativistic scheme would appear the sum of the two intervening
potentials in the Schr\"{o}dinger equation and no bound-state solutions
would be possible for $\lambda _{r}<0$ and $|\lambda _{r}|>|\lambda _{0}|$.
Therefore, the weak-coupling limit does not correspond to the
nonrelativistic limit and so we can say that the nonminimal vector linear
potential given by (\ref{pot2}) is an intrinsically relativistic potential
in the DKP theory.

\section{Conclusions}

We showed that minimal and nonminimal vector interactions behave differently
under the charge-conjugation transformation. In particular, nonminimal
vector interactions have no counterparts in the Klein-Gordon theory. The
conserved charge current plus the charge conjugation operation are enough to
infer about the absence of Klein's paradox under nonminimal vector
interactions, or its possible presence under minimal vector interactions.
Although Klein's paradox can not be treated as unworthy of regard in the DKP
theory with minimally coupled vector interactions, it never makes its
appearance in the case of nonminimal vector interactions because they do not
couple to the charge. Nonminimal vector interactions have the very same
effects on both particles and antiparticles and so in the case of a pure
nonminimal vector coupling, both particle and particle energy levels are
members of the spectrum, and the particle and antiparticle spectra are
symmetrical about $E=0$. If the interaction potential is attractive
(repulsive) for bosons it will also be attractive (repulsive) for
antibosons. However, there is no crossing of levels because possible states
in the strong field regime with $E=0$ are in fact unnormalizable. These
facts imply that there is no channel for spontaneous boson-antiboson
creation and for that reason the single-particle interpretation of the DKP
equation is ensured. The charge conjugation operation allows us to migrate
from the spectrum of particles to the spectrum of antiparticles and vice
versa just by changing the sign of $E$. This change induces no change in the
nodal structure of the components of the DKP spinor and so the nodal
structure of the four-current is preserved.

We showed that nonminimal vector couplings have been used improperly in the
phenomenological description of elastic meson-nucleus scatterings potential
by observing that the four-current is not conserved when one uses either the
matrix $P\beta ^{\mu }$ or $\beta ^{\mu }P$, even though the bilinear forms
constructed from those matrices behave as true Lorentz vectors. The space
component of the nonminimal vector potential can not be absorbed into the
spinor and we showed that the space component of the nonminimal vector
potential could be irrelevant for the formation of bound states for
potentials vanishing at infinity but its presence is an essential ingredient
for confinement.

The complete solution of the DKP equation with spherically symmetric
nonminimal vector potentials was found by recurring to vector spherical
harmonics due to the expression appearing in (\ref{dkp3}) with $%
\overrightarrow{\nabla }$ in spherical coordinates acting on a function of $%
r $ multiplied by $Y_{lm_{l}}(\theta ,\varphi )$. A similar procedure
resulting in a set of coupled differential equations for the components of
the spinor has already appeared in the literature \cite{ned1}. Here, instead
of a set of coupled first-order equations, the DKP equation was mapped into
a Sturm-Liouville problem for the first component of the spinor and the
remaining components were expressed in terms of the first one in a simple
way. In this process, the conserved four-current was also expressed in terms
of the first component of the DKP spinor in such a way that the searching
for the solutions of the DKP equation becomes more clear and transparent.
The conservation of the total angular momentum was derived from its
commutation properties with each term of the DKP equation.

The solution for a nonminimal linear potential was found by solving a Schr%
\"{o}dinger-like problem for the nonrelativistic harmonic oscillator for the
first component of the spinor. The behavior of the solutions for this sort
of DKP oscillator was discussed in detail. Instead of imposing boundary
conditions at the origin by recurring to plausibility arguments regarding
the self-adjointness of the momentum and the finiteness of the kinetic
energy, as done by Greiner \cite{gre} in the case of the nonrelativistic
harmonic oscillator, the proper boundary conditions were imposed in a simple
way by observing the absence of Dirac delta potentials. The exact solutions
were presented in a closed form and the spectrum presents, beyond the
essential degeneracy omnipresent for any central force field, an accidental
degeneracy. That model reinforced the absence of Klein's paradox for
nonminimal vector interactions.

\vspace{2.5cm}

\noindent \textbf{Acknowledgments}

We acknowledge financial support from CAPES and CNPq.

\end{document}